\documentclass[sigplan,screen,nonacm,10pt]{acmart}
\pdfoutput=1

\usepackage{url}
\usepackage{booktabs}
\usepackage{array}
\usepackage{graphicx}
\usepackage{amsmath}
\usepackage{ifthen}
\usepackage{multirow}
\usepackage{multicol}
\usepackage{enumitem}
\usepackage{xspace}
\usepackage{xcolor}
\usepackage{balance}
\usepackage[linesnumbered, ruled]{algorithm2e}

\SetKwFor{For}{for}{do}{}
\SetKwFor{ForEach}{foreach}{do}{}
\SetKwIF{If}{ElseIf}{Else}{if}{}{elif}{else}{}
\SetKwFor{While}{while}{do}{}
\SetKwRepeat{Repeat}{repeat}{until}
\SetKwRepeat{Do}{do}{while}
\SetKwInput{Input}{Input}
\SetKwInput{Output}{Output}
\SetKw{Continue}{continue}

\newcommand{\xxx}{\textsc{Themis}\xspace}
\newcommand{\yyy}{FSSDP\xspace}
\newcommand{\YYY}{Fully Sharded Sparse Data Parallelism\xspace}

\newcommand{\mqwen}{Qwen1.5-MoE\xspace}
\newcommand{\mphi}{Phi-3.5-MoE\xspace}
\newcommand{\mds}{DeepSeek-V3\xspace}

\newcommand{\clustera}{Cluster A\xspace}
\newcommand{\clusterh}{Cluster B\xspace}

\newcommand{\ie}{i.e.\xspace}
\newcommand{\eg}{e.g.\xspace}

\newcommand{\refsec}[1]{\S~\ref{#1}}
\newcommand{\reffig}[2][]{Figure~\ref{#2}\ifthenelse{\equal{#1}{}}{}{#1}}

\newcommand{\refalgo}[1]{Algorithm~\ref{#1}}
\newcommand{\refeq}[1]{Equation~\ref{#1}}

\newenvironment{niceritemize}{%
    \begin{itemize}[left=0pt, topsep=0pt, partopsep=0pt, parsep=0pt, itemsep=0pt]
}{%
    \end{itemize}
}

\newcommand{\set}[1]{\mathcal{#1}}
\newcommand{\gset}[1]{\set{#1}^\text{g}}
\newcommand{\setE}{\mathcal{E}}

\newcommand{\coll}[1]{\texttt{#1}\xspace}
\newcommand{\collar}{\coll{All\-Reduce}}
\newcommand{\collatoa}{\coll{All-to-All}}
\newcommand{\collbc}{\coll{Broadcast}}
\newcommand{\collrd}{\coll{Reduce}}
\newcommand{\collag}{\coll{All\-Gather}}
\newcommand{\collrs}{\coll{Reduce\-Scatter}}
\newcommand{\collsag}{\coll{Sparse\-All\-Gather}}
\newcommand{\collsrs}{\coll{Sparse\-Reduce\-Scatter}}
\newcommand{\collsaga}{\coll{SpAG}}
\newcommand{\collsrsa}{\coll{SpRS}}

\newcommand{\collsagop}[2]{\mbox{\texttt{SpAG(}$#1$, $#2$\texttt{)}}}
\newcommand{\collsrsop}[2]{\mbox{\texttt{SpRS(}$#1$, $#2$\texttt{)}}}

\newcommand{\bw}{\text{bw}}

\newcommand{\imbalanceSlowdown}{2.87$\times$\xspace}
\newcommand{\memorySpeedup}{2.65$\times$\xspace}
\newcommand{\timelinessSpeedup}{2.9\%\xspace}
\newcommand{\timelinessSlowdown}{10.2\%\xspace}

\newcommand{\numclusters}{2\xspace}
\newcommand{\overallSpeedup}{1.26-2.42$\times$\xspace}
\newcommand{\adamMemFactor}{6$\times$\xspace}
\newcommand{\rearrangeReserveFactor}{4$\times$\xspace}
\newcommand{\rearrangeFreqA}{25\xspace}
\newcommand{\rearrangeFreqB}{10\xspace}

\newcommand{\maxSpeedupEP}{3.68$\times$\xspace}
\newcommand{\avgSpeedupA}{2.41$\times$\xspace}
\newcommand{\avgSpeedupH}{2.60$\times$\xspace}
\newcommand{\minSpeedupFasterMoE}{1.26$\times$\xspace}
\newcommand{\maxSpeedupFasterMoE}{2.37$\times$\xspace}
\newcommand{\maxSpeedupSmartMoE}{3.27$\times$\xspace}
\newcommand{\maxSpeedupFlexMoE}{2.92$\times$\xspace}
\newcommand{\speedupUniformMin}{1.07$\times$\xspace}
\newcommand{\speedupUniformMax}{2.27$\times$\xspace}
\newcommand{\imbalanceOverheadMin}{1.48$\times$\xspace}
\newcommand{\imbalanceOverheadMax}{2.87$\times$\xspace}

\newcommand{\layerwiseSpeedupRange}{2.1-8.1$\times$\xspace}
\newcommand{\ataReduction}{13.0$\times$\xspace}
\newcommand{\rmOverheadIncrease}{33.3\%\xspace}
\newcommand{\rmSpeedupOverBaselines}{1.27$\times$\xspace}

\newcommand{\flexmoeMemOverhead}{3.0$\times$\xspace}
\newcommand{\themisParamMemVsEP}{3.1$\times$\xspace}
\newcommand{\themisTotalMemIncrease}{35.6\%\xspace}
\newcommand{\rmParamMemReduction}{45.4\%\xspace}
\newcommand{\rmTotalMemIncrease}{17.8\%\xspace}

\newcommand{\heteroSpeedupA}{1.05$\times$\xspace}
\newcommand{\heteroSpeedupB}{1.03$\times$\xspace}
\newcommand{\heteroSpeedupC}{1.08$\times$\xspace}
\newcommand{\heteroSpeedupD}{1.08$\times$\xspace}
\newcommand{\matHeteroVsHetero}{2.88$\times$\xspace}
\newcommand{\matHeteroVsMat}{1.11$\times$\xspace}
\newcommand{\scalabilitySpeedupMin}{1.48$\times$\xspace}
\newcommand{\scalabilitySpeedupMax}{3.35$\times$\xspace}

\newcommand{\loadPredModel}{\mqwen}
\newcommand{\loadPredCluster}{\clustera}
\newcommand{\loadPredRSquared}{0.99\xspace}
 
\begin{document}

\title[Themis: Efficient Sparse Model Training Through FSSDP]{Themis: Efficient Sparse Model Training\texorpdfstring{\\}{ }Through Fully Sharded Sparse Data Parallelism}

\author{Yuhao Qing}
\authornote{Yuhao Qing and Guichao Zhu contributed equally to this work.}
\affiliation{%
  \institution{The University of Hong Kong}
  \country{}}
\email{qyhh@connect.hku.hk}

\author{Guichao Zhu}
\authornotemark[1]
\affiliation{%
  \institution{The University of Hong Kong}
  \country{}}
\email{gczhu@connect.hku.hk}

\author{Lintian Lei}
\affiliation{%
  \institution{The University of Hong Kong}
  \country{}}
\email{leilt@connect.hku.hk}

\author{Fanxin Li}
\affiliation{%
  \institution{The University of Hong Kong}
  \country{}}
\email{fxli@connect.hku.hk}

\author{Shixiong Zhao}
\affiliation{%
  \institution{The University of Hong Kong}
  \country{}}
\email{zsxhku@connect.hku.hk}

\author{Zekai Sun}
\affiliation{%
  \institution{The University of Hong Kong}
  \country{}}
\email{zksun@cs.hku.hk}

\author{Xiuxian Guan}
\affiliation{%
  \institution{The University of Hong Kong}
  \country{}}
\email{guanxx@connect.hku.hk}

\author{Xusheng Chen}
\affiliation{%
  \institution{The University of Hong Kong}
  \country{}}
\email{chenxus@connect.hku.hk}

\author{Dong Huang}
\affiliation{%
  \institution{The University of Hong Kong}
  \country{}}
\email{dhuang@cs.hku.hk}

\author{Ping Luo}
\affiliation{%
  \institution{The University of Hong Kong}
  \country{}}
\email{pluo.lhi@gmail.com}

\author{Yiming Qiu}
\affiliation{%
  \institution{The University of Hong Kong}
  \country{}}
\email{yimingq@hku.hk}

\author{Heming Cui}
\affiliation{%
  \institution{The University of Hong Kong}
  \country{}}
\email{heming@cs.hku.hk}

\renewcommand{\shortauthors}{Qing et al.}

\begin{abstract}
Mixture-of-Experts (MoE) scales large language models cost-effectively, but expert-parallel training suffers severe straggler effects from skewed expert loads.
Current systems frequently rearrange expert placement to mitigate stragglers, inflating memory footprint and migration overhead, potentially negating the benefits of load balancing.
We present \textit{\YYY (\yyy)}, a sparse-native MoE training approach that enables in-situ load balancing on every training iteration, overlapping the balancing with computation and eliminating explicit expert rearrangement together with its migration traffic and memory reserves.
\yyy keeps MoE layers sharded and sparsely materializes an ephemeral, load-balancing placement each iteration, with re-materialization to reuse available memory across layers.
\yyy is complemented by heterogeneous sharding to shift memory imbalance from the device level to the layer level, maintaining uniform memory budgets while enabling per-layer placement optimization.
We realize \yyy in \xxx with co-designed topology-aware placement algorithms.
Across \numclusters clusters and diverse workloads, \xxx achieves \overallSpeedup speedup over state-of-the-art expert-rearrangement systems. \end{abstract}


\maketitle

\section{Introduction}\label{sec:intro}

Large language models (LLMs) increasingly adopt Mixture-of-Experts (MoE) to scale cost-effectively~\cite{jacobs1991adaptive,shazeer2017outrageously,fedus2022switch,jiang2024mixtral,deepseekai2025deepseekv3technicalreport,kimiteam2025kimik2openagentic} by decomposing dense layers into many sub-networks called \textit{experts} and conditionally activating only a small subset per input.
MoE training typically employs \textit{expert parallelism} (EP)~\cite{shazeer2017outrageously}, which evenly distributes experts across devices (\eg, GPUs) and dispatches inputs via \collatoa under a jointly trained routing network (\ie, \textit{gate}).
However, due to the static expert assignment on devices, as the gate evolves with training, skewed routing induces load imbalance across devices, where some receive far more inputs than others, creating hotspots on both devices and interconnects and ultimately causing up to a \imbalanceSlowdown slowdown in throughput compared to balanced routing (\refsec{sec:eval:end2end}).

To mitigate stragglers, existing systems~\cite{nie2023flexmoe,zhai2023smartmoe,he2022fastermoe} resort to \textit{expert rearrangement}, which adaptively rearranges \textit{expert placement} (\ie, the presence of each expert on each device) by relocating or replicating experts across devices, as depicted in \reffig[d]{fig:overlap}.
For every replicated expert, its replicas' gradients must be synchronized after each iteration.
By rebalancing device workloads under a new placement, these systems strive to mitigate stragglers in the iterations between frequent rearrangements.

\begin{figure*}[t!]
    \centering

    \includegraphics[width=\textwidth]{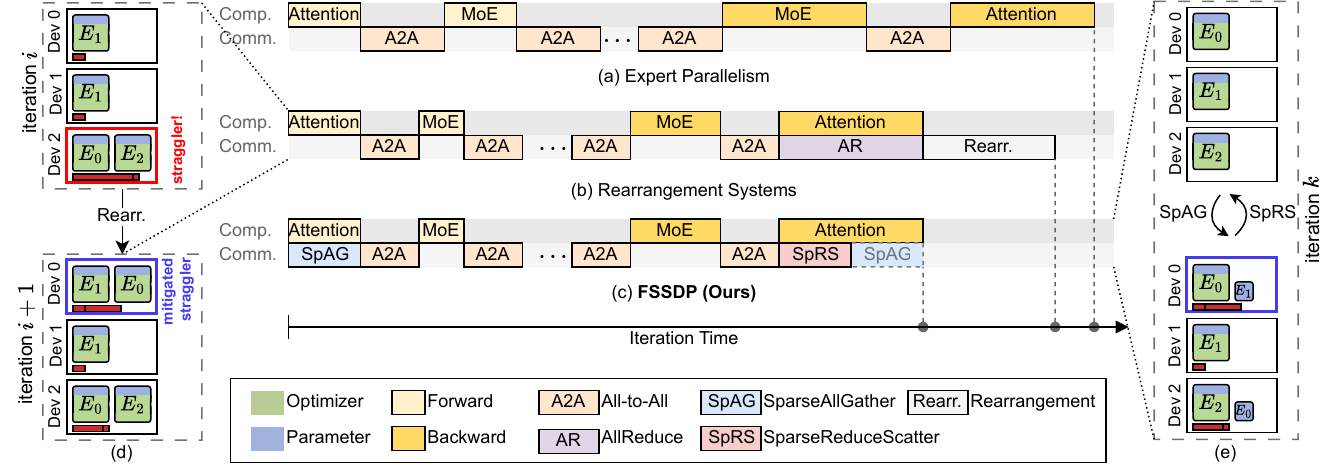}

    \caption{
        MoE parallel training strategies for a single Transformer layer with MoE. 
        The ellipses between the \collatoa operations represent other layers in the model. 
        (a) EP suffers stragglers on overloaded devices.
        (d) Rearrangement adjusts expert placement to mitigate straggler effects of EP. The red bars under experts show per-device workloads. 
        (b) Rearrangement systems have lower MoE computation and \collatoa latency than EP, but introduce rearrangement overhead on the critical path and \collar for replica gradient synchronization.
        (e) Using two sparse collectives, \yyy achieves in every iteration the same balanced placement as rearrangement, while avoiding rearrangement overhead between iterations, as shown in (c). The dashed-line \collsag box re-materializes released parameters for the preceding MoE layer's backward computation if needed.
    }

    \label{fig:overlap}
\end{figure*}
 
Despite these efforts, rearrangement introduces two fundamental system challenges.
First, \textbf{memory constraints limit placement flexibility (C1)}.
A placement that better balances load can be more memory-hungry, as expert replication multiplies memory consumption, while expert colocation concentrates memory usage on certain devices, skewing memory usage across devices.
Hosting an incoming expert persistently requires reserving memory for its parameters and their larger optimizer states (e.g., \adamMemFactor for Adam~\cite{kingma2014adam} in mixed precision training).
Limited reserves constrain the system to suboptimal placements, while excessive reserves starve the training workload.
Our experiment on FlexMoE~\cite{nie2023flexmoe} shows that achieving a \memorySpeedup speedup requires \rearrangeReserveFactor more reserved memory for rearrangement.

Second, \textbf{rearrangement frequency trades timeliness against communication overhead (C2)}.
Adapting to shifting loads demands frequent rearrangements, but each adds critical-path migration communication, making frequent rearrangement counterproductive.
Our experiment on SmartMoE~\cite{zhai2023smartmoe} shows that increasing frequency (\eg, from every \rearrangeFreqA steps to \rearrangeFreqB steps) shortens non-rearrangement iterations (\eg, by \timelinessSpeedup) yet increases end-to-end time (\eg, by \timelinessSlowdown) from accumulated migration overhead.

We argue that the root cause is \textit{the statefulness of expert placement}. Replicated parameters and optimizer states persist between rearrangements, while routing behavior evolves at the iteration timescale, creating a misalignment.
To support load-balancing placement persistently over iterations, rearrangement moves expert state across devices, which reintroduces the memory pressure that EP sought to eliminate.
Treating placement as persistent state forces systems into perpetual migration to chase the gate's shifting patterns.
Consequently, expert replication and colocation inflate memory footprint as well as skew memory distribution across devices (C1), and frequent state migration injects communication overhead directly into the critical path (C2).

In this work, our goal is the full load-balancing benefit of flexible placement without rearrangement's memory reserves or critical-path migration.
We seek to minimize the reliance of the load balancer on persistent-state migration, so a system can respond to routing dynamics.
Rather than deciding placement and migration expert by expert, we take a holistic, layer-centric perspective that treats the MoE FFN as a unified sparse layer, enabling ephemeral per-iteration placement materialization.
Through rethinking MoE-style sparsification on dense models (detailed in \refsec{sec:bg}), we achieve this goal based on two insights, which together yield the throughput and memory-resilience gains measured in \refsec{sec:eval}.

Our first insight is that the extra replicas a placement introduces can be materialized ephemerally each iteration at the same total communication volume that rearrangement-based systems already pay to synchronize replica gradients across those replicas (\refsec{sec:design:cost}).
We propose \collsag and \collsrs, sparse collectives for this ephemeral parameter materialization and gradient reduction.
Placement thus becomes stateless, as adopting a new placement each iteration migrates no optimizer state.

Our second insight is that placement flexibility and memory balance, seemingly contradictory goals, can be reconciled through cross-layer redistribution.
Flexible expert placement within a layer inherently creates memory imbalance, as devices hosting more experts for load balancing consume more memory.
By inversely distributing memory across layers (large shards in layer $i$ paired with small shards in layer $j$ on the same device), we shift heterogeneity from the device level to the layer level, maintaining uniform total memory while enabling per-layer placement optimization.

We present \textit{\YYY (\yyy)}, a sparse-native MoE training approach that enables in-situ expert load balancing in each training iteration, eliminating explicit expert rearrangement and the persistent-state migration it requires.
\yyy keeps each expert's optimizer state complete on a single \textit{owner} device, with no redundant replicas.
In each iteration, \yyy \textit{sparsely materializes} MoE parameters for an ephemeral load-balancing expert placement (see \reffig[c]{fig:overlap}).
To reuse memory across layers, \yyy optionally releases parameters after use and \textit{re-materializes} them in the backward pass.
Through \textit{heterogeneous sharding}, \yyy redistributes memory across layers to enable per-layer placement optimization while maintaining uniform per-device memory budgets.

We build \xxx, a high-throughput distributed MoE training engine on top of \yyy, improving the memory and communication efficiency of expert load balancing over rearrangement-based baselines.
\xxx employs topology-aware algorithms that optimize \yyy for network hierarchies and overlap sparse collectives with computation, leaving only a small residual communication cost on the critical path (\refsec{sec:eval}).
Our evaluations across \numclusters clusters and diverse workloads show that \xxx achieves \overallSpeedup speedup over the state-of-the-art expert-rearrangement systems FasterMoE, SmartMoE, and FlexMoE~\cite{he2022fastermoe,zhai2023smartmoe,nie2023flexmoe}.
The improvements are more pronounced under higher routing skew, larger parallel scale, and tighter memory budgets (\refsec{sec:eval}).

The main contributions of this work are as follows:
\begin{niceritemize}
    \item We identify stateful expert placement as the root cause of rearrangement's memory and critical-path overheads (C1, C2), and propose \yyy, which keeps each expert's parameters and optimizer states complete in a single shard and sparsely materializes each iteration's placement as ephemeral replicas beyond those shards.
    \item We design heterogeneous sharding and re-materialization, which redistribute and reuse memory across layers, respectively. Together, they reconcile placement flexibility with uniform per-device memory budgets.
    \item We develop \xxx, a high-performance MoE training engine that combines \yyy with topology-aware expert placement scheduling strategies.
    \item We evaluate \xxx across diverse hardware and workloads and show consistent improvements over baselines in all evaluated configurations.
\end{niceritemize}
\section{Background and Motivation}\label{sec:bg}

We review existing strategies for scaling feed-forward networks and present a new perspective, viewing MoE as decomposing wide networks into smaller sub-networks rather than composing replicated components.
We then analyze how current approaches lead to designs that perpetually chase shifting loads through state movement.

\subsection{Scaling FFNs with Parallelism}

LLMs heavily rely on stacked Transformer layers.
Each layer typically includes a Feed-Forward Network (FFN), constituting a significant portion of the layer's parameters and computation.
A typical FFN processes input tokens $x$ through two linear transformations with activation $f$, defined in \refeq{eq:ffn} (biases omitted for brevity).
\begin{equation}\label{eq:ffn}
    \text{FFN}(x) = f(x \cdot W^{A}) \cdot W^{B}
\end{equation}
As model scale increases~\cite{kaplan2020scaling,hoffmann2022chinchilla}, these dense FFNs exceed a single device's memory and compute capacity, necessitating distributed training.

\textbf{Data Parallelism (DP)}~\cite{li2020pytorchddp} is a common parallelization technique that replicates the entire model on each device and processes different data partitions in parallel, synchronizing gradients via \collar.
However, for a large FFN, replicating the full parameters and optimizer states on every device is often infeasible due to memory constraints.

\textbf{Tensor Parallelism (TP)}~\cite{shoeybi2019megatron,Korthikanti2022ReducingAR} partitions the weight matrices across devices, so each device computes on its local slice, as shown in \refeq{eq:ffn_tp} and \reffig[c]{fig:ffn_parallelism}.
\begin{align}\label{eq:ffn_tp}
    \text{FFN}_{\text{TP}}(x) &= f(x \cdot [W_1^A, \ldots, W_n^A]) \cdot [W_1^B, \ldots, W_n^B]^T \nonumber\\
    &= \sum_{i=1}^{n} \underbrace{f(x \cdot W_{i}^{A}) \cdot W_{i}^{B}}_{\text{TP partition computation}}
\end{align}
However, TP requires collective communication (\eg, \collag, \collrs) involving \textit{all} tokens across \textit{all} devices within the TP group, incurring significant communication overhead.

\begin{figure*}[t!]
    \centering
    \includegraphics[width=\textwidth]{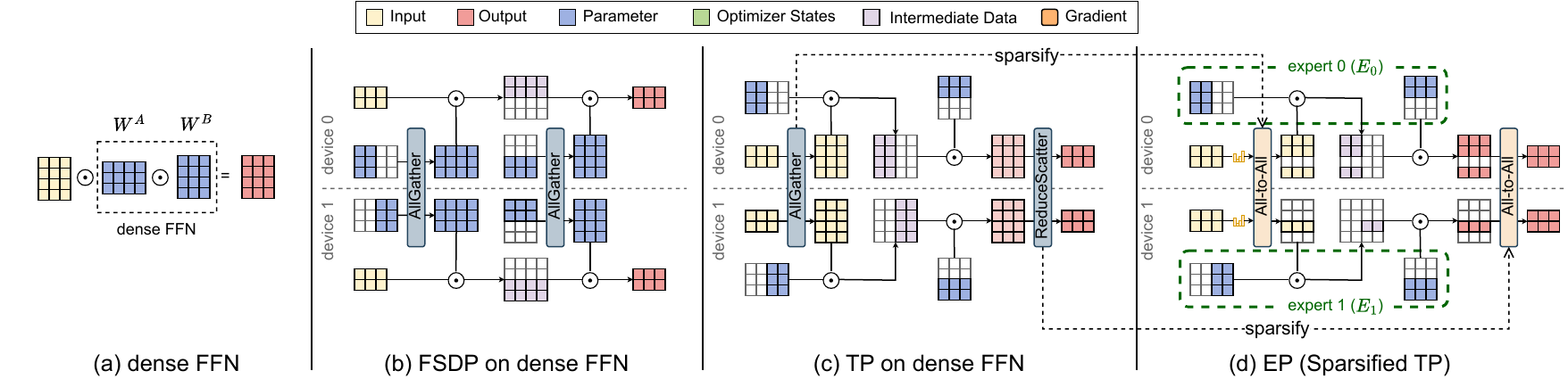}
    \caption{
        From dense FFN parallelization (FSDP, TP) to sparse MoE with EP and conditional computation.
    }
    \label{fig:ffn_parallelism}
\end{figure*}
 
\textbf{Fully Sharded Data Parallelism (FSDP)}~\cite{rajbhandari2020zero,zhao2023pytorchfsdp} shards all model parameters, gradients, and optimizer states across data-parallel devices (\reffig[b]{fig:ffn_parallelism}) for memory efficiency.
During computation, each device gathers the parameters for its local data batch using \collag, computes, and then discards the parameters.
Gradients are reduced using \collrs after the backward pass.

\subsection{MoE Sparsification and Expert Parallelism}
Instead of computing the dense FFN, MoE approximates it by \textit{sparsifying} the FFN with conditional computation.
MoE can be viewed as decomposing a wide FFN into smaller sub-networks called \textit{experts} ($E_i$), where each expert corresponds to the partitioned blocks ($W_i^A$, $W_i^B$) from TP (\refeq{eq:ffn_tp}).
A lightweight gating network $g(\cdot)$ selects a small subset (\eg, top-$k$) of these experts for each input token, approximating the dense FFN's output through their weighted sum (\refeq{eq:moe}).
Only selected experts process that token, reducing computation relative to the dense FFN.
\begin{equation}\label{eq:moe}
    \text{MoE}(x) = \sum_{i \in \arg\text{topk}(g(x))} g(x)_i \cdot \underbrace{f(x \cdot W_i^{A}) \cdot W_i^{B}}_{\text{Expert } E_i \text{ computation}}
\end{equation}

\textbf{Expert Parallelism (EP)}~\cite{shazeer2017outrageously} can be viewed as sparse tensor parallelism that distributes experts (analogous to TP partitions) across devices (\reffig[d]{fig:ffn_parallelism}) but prunes TP's dense communication to \collatoa, routing each token only to devices holding its assigned experts.

\begin{figure}[h!tb]
    \centering
    \includegraphics[width=.8\linewidth]{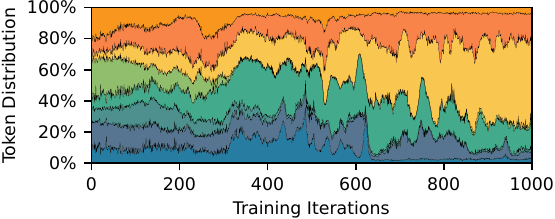}

    \caption{
        Skewed and fluctuating expert workload in GPT-S-MoE training with 8 experts under the auxiliary load-balancing loss (weight 0.001).
        Each band is one expert; band height shows its share of routed tokens.
    }

    \label{fig:trace}
\end{figure}

\textbf{Straggler Effects.}
EP suffers from critical straggler effects in practice.
The trainable gate evolves during training, producing \textit{fluctuating} and \textit{skewed} load distributions~\cite{gale2023megablocks,hwang2022tutel,nie2023flexmoe,li2023lina,he2022fastermoe,zhai2023smartmoe,fedus2022switch}.
Some experts receive disproportionately many tokens, with loads shifting dramatically across iterations (\reffig{fig:trace}).
This skewness creates stragglers on devices hosting highly-routed experts, degrading training throughput and eroding MoE's sparsity benefits.
Routing-side remedies temper the skew but do not eliminate the drift behind it.
Even raising the auxiliary load-balancing loss weight~\cite{lepikhin2020gshard} to 0.1 leaves substantial residual imbalance in our measurements (\reffig[a]{fig:sensitivity}).
This motivates system-level load balancing that acts on expert placement, complementing these routing-side remedies without changing the training algorithm.
 
\subsection{Expert Rearrangement}
To mitigate stragglers, recent systems~\cite{he2022fastermoe,zhai2023smartmoe,nie2023flexmoe} dynamically rearrange expert placements across devices to redistribute experts based on their load.
Overloaded experts can be replicated across devices to spread incoming tokens, while underloaded experts can be colocated to consolidate loads and free memory for replicas.
This rebalancing transforms the device-level workload distribution, as illustrated in \reffig[a]{fig:ffn_placement} and \reffig[d]{fig:overlap}.
SmartMoE~\cite{zhai2023smartmoe} exchanges experts between placement slots, moving parameters together with their optimizer states; FlexMoE~\cite{nie2023flexmoe} expands, shrinks, and migrates replicas, copying optimizer states on expansion; FasterMoE~\cite{he2022fastermoe} broadcasts overloaded experts' weights to all devices once the iteration's routing is known.

\begin{figure*}[t!]
    \centering
    \includegraphics[width=\textwidth]{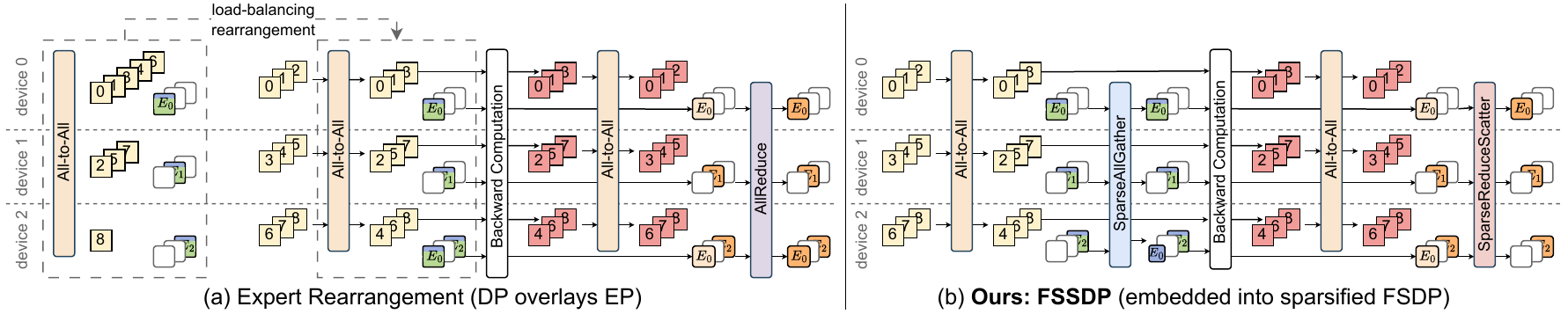}
    \caption{
        Rearrangement-based approaches and \yyy support the same expert placement in distinct ways.
    }
    \label{fig:ffn_placement}
\end{figure*}
 
However, rearrangement faces significant challenges.
Rearrangement must reserve memory for an incoming expert's parameters and optimizer states, totaling 7$\times$ the expert's half-precision parameter size in mixed precision training~\cite{micikevicius2017mixed} with Adam~\cite{kingma2014adam}, whose FP32 master weights and two moment terms add 6$\times$ beyond the 1$\times$ parameters, with extra space for gradients and transfer staging.
Persistent expert placement also skews memory distribution across devices. Colocating $k$ underloaded experts and replicating overloaded ones across devices leaves some devices underutilized and others oversubscribed, constraining placement flexibility.

Rearrangement also incurs substantial communication overhead on the critical path.
For instance, migrating one expert per layer of \mphi~\cite{abdin2024phi3technicalreporthighly} requires transferring 5\,GB of parameters and 30.2\,GB of optimizer states, taking 0.7 seconds even on a 400\,Gbps interconnect.
To bound these overheads, FasterMoE~\cite{he2022fastermoe} replicates only experts whose modeled benefit exceeds the broadcast and reduction cost, while SmartMoE~\cite{zhai2023smartmoe} and FlexMoE~\cite{nie2023flexmoe} leave rearrangement frequency to user tuning.
Memory reserves and migration cost thus constrain how aggressively rearrangement can chase shifting loads.

\subsection{Rethinking MoE Training Parallelization}
Existing approaches graft DP-style replication or expert relocation atop EP, an expert-centric design where this balancing layer continually chases stragglers caused by EP and by routing dynamics that shift at the iteration timescale.
This misalignment calls for unifying sparsity and load balancing at the core of the parallelization design.
A natural question is whether switching to another parallelism, such as FSDP, to parallelize the MoE FFN would suffice.
Unfortunately, FSDP treats the layer as dense, collapsing sparsity and forcing every step to \collag and \collrs all experts, so communication scales with the total number of experts rather than the number activated.
By shifting our perspective from sparsifying TP to sparsifying FSDP, we keep the ``shard and materialize'' pattern but apply it sparsely. Optimizer states remain sharded at expert granularity, complete on one device each, and only the expert replicas that the iteration's placement adds beyond those shards are materialized, with their gradients later reduced back.
Because shard boundaries coincide with expert boundaries, each expert executes in place at its shard without any parameter movement. Per-iteration parameter movement thus scales with the placement delta rather than with the number of executing experts.
This yields \yyy, a layer-centric, unified parallelization approach that anchors each expert's parameters and optimizer states in a single shard and embeds per-iteration ephemeral expert placement on top, as depicted in \reffig[b]{fig:ffn_placement}.
\section{\YYY}\label{sec:design}

\begin{figure*}[t!]
    \centering

    \includegraphics[width=\textwidth]{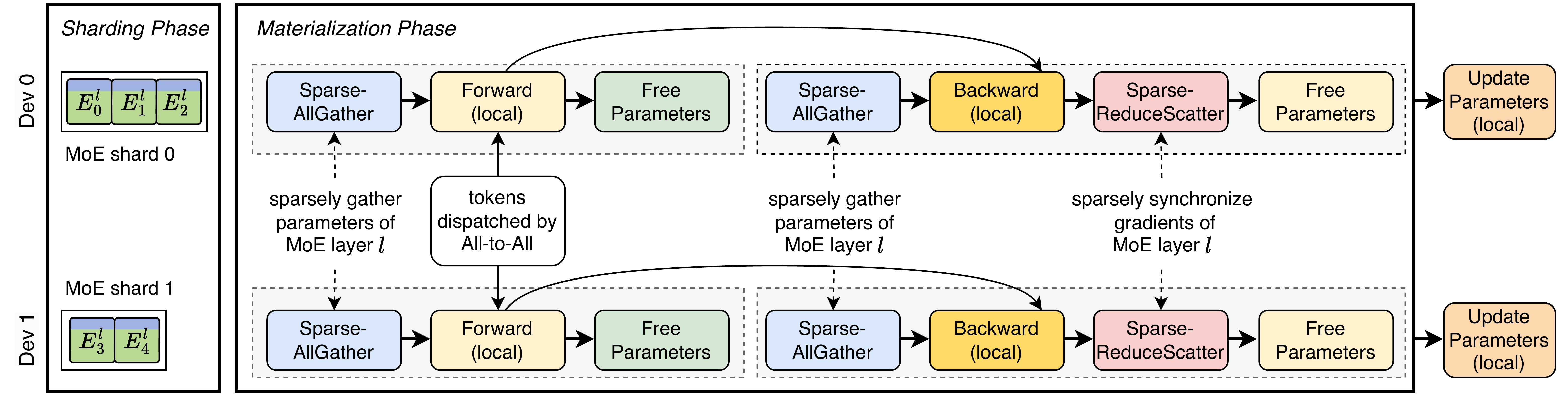}

    \caption{
        Workflow of \yyy at MoE layer $l$ in an iteration.
        $E^{l}_{i}$ represents expert $i$ of MoE layer $l$ in the LLM.
        The \textit{sharding phase} partitions the MoE layer's parameters and optimizer states into MoE shards placed across devices.
        In the \textit{materialization phase}, the forward and backward passes execute with two sparse collectives, \collsag and \collsrs.
        With re-materialization enabled, materialized parameters are freed after each use, and a second \collsag runs before the backward pass.
    }

    \label{fig:workflow}
\end{figure*}
 
\YYY (\yyy) comprises two phases, as depicted in~\reffig{fig:workflow}: (1) the \textit{sharding phase}, where parameters and optimizer states are partitioned into MoE shards across devices; and (2) the \textit{materialization phase}, where an ephemeral expert placement is materialized using two sparse communication collectives, \collsag and \collsrs.
We first describe the two phases (\refsec{sec:design:model}), then define the two sparse collectives (\refsec{sec:design:spcoll}), and analyze their communication cost against FSDP-style and rearrangement-based alternatives (\refsec{sec:design:cost}).

\subsection{MoE Sharding and Ephemeral Placement}\label{sec:design:model}

\yyy splits an MoE layer's parameters, and likewise their gradients, into equal-sized chunks $\set{C} = \{C_0, C_1, \ldots\}$ with one chunk per expert.
Let $\set{D} = \{D_0, D_1, \ldots\}$ denote all devices in the communication group.
A chunk placement $\set{P} \subseteq \set{C} \times \set{D}$ specifies data distribution, where $(c, d) \in \set{P}$ indicates that chunk $c$ resides on device $d$.
A \textit{shard placement} assigns each chunk to exactly one device, while a device may hold multiple chunks.

During the sharding phase, each MoE layer is partitioned into $|\set{D}|$ disjoint \textit{MoE shards}, with experts as atomic units.
Each shard, uniquely assigned to one device, contains parameters and optimizer states of a subset of experts.
The sharding phase thus fixes the shard placement $\set{P}$ of each MoE layer, and each device applies the optimizer updates of the experts it owns.
Shard sizes need not be equal across devices, as \yyy requires only a shard placement.

In the materialization phase, each training iteration executes the MoE layer under its own ephemeral load-balancing placement $\set{P}' \supseteq \set{P}$.
In the forward pass, \collsag (\refsec{sec:design:spcoll}) materializes the expert replicas that $\set{P}'$ adds beyond the shards, forming $\Delta\set{P} = \set{P}' \setminus \set{P}$, while experts already in $\set{P}$ execute in place without parameter movement.
Tokens are then dispatched among the devices holding their assigned experts.
In the backward pass, \collsrs reduces each expert's gradients from all devices holding its replicas back to the device hosting the corresponding MoE shard, which then updates parameters and optimizer states locally.
Ephemeral replicas carry no optimizer state.
To reuse memory across layers, re-materialization discards materialized MoE parameters after use, at the cost of one additional \collsag toward the same placement $\set{P}'$ per MoE layer per iteration.
Released replicas do not accumulate across layers, so the reclaimed memory either lowers the peak memory footprint or allows each layer more replicas within the same memory.

\subsection{Sparse Collectives}\label{sec:design:spcoll}

Each sparse collective is defined by two chunk placements, a pre-condition and a post-condition.

\begin{sloppypar}
\textbf{\collsag} (\collsaga) specifies the pattern of data movement for materializing expert parameters.
Its pre-condition is the shard placement $\set{P}$; \collsaga selectively materializes the added replicas $\Delta\set{P}$, yielding the iteration's placement $\set{P}'$ as the post-condition.
Thus, a specific \collsag can be formulated as:
\begin{equation*}
    \collsagop{\set{P}}{\set{P}'}
    \;\;
    \text{s.t. } \set{P} \text{ is a shard placement and } \set{P} \subseteq \set{P}'
\end{equation*}
\end{sloppypar}

\begin{figure}[btp]
    \centering
    \includegraphics[width=\linewidth]{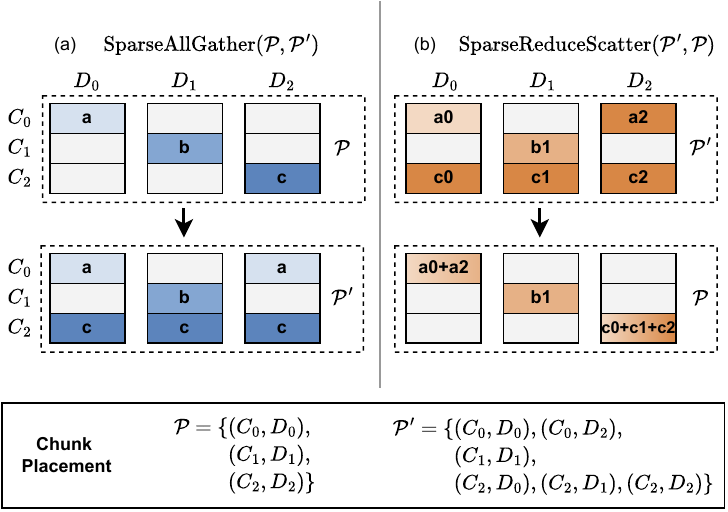}

    \caption{
        An example of the two symmetric sparse collectives, \collsag and \collsrs.
        (a) \collsag delivers exactly the (chunk, device) pairs that the post-condition $\set{P}'$ adds beyond the shard placement $\set{P}$, where each chunk resides on exactly one device in $\set{P}$; pairs already in $\set{P}$ move nothing.
        (b) \collsrs sums each chunk's gradients over its copies in $\set{P}'$ and places the result at the chunk's shard in $\set{P}$.
    }

    \label{fig:spcoll}
\end{figure}

\textbf{\collsrs} (\collsrsa) reduces gradients from the ephemeral expert replicas.
Each chunk in the post-condition, the shard placement $\set{P}$, is the sum of that chunk's gradients across all devices holding it in the pre-condition $\set{P}'$.
A specific \collsrs can be formulated as:
\begin{equation*}
    \collsrsop{\set{P}'}{\set{P}}
    \;\;
    \text{s.t. } \set{P} \text{ is a shard placement and } \set{P} \subseteq \set{P}'
\end{equation*}
Within one iteration, \collsagop{\set{P}}{\set{P}'} is thus paired with the symmetric \collsrsop{\set{P}'}{\set{P}}, as exemplified in \reffig{fig:spcoll}.

\subsection{Communication Cost Analysis}\label{sec:design:cost}

We now compare \yyy's per-iteration state movement against three alternatives: FSDP, selective expert-wise FSDP, and rearrangement.
One cost model covers all three comparisons, charging each approach the bytes it communicates and the time those transfers take.
With an MoE layer of size $S$ split into $|\set{C}|$ equal chunks, each chunk carries $\Psi = S/|\set{C}|$ bytes, the size of one expert.
The model counts $\Psi$ bytes when a materialized replica enters its destination device and $\Psi$ bytes when that replica's gradient leaves the same device, once per replica in each direction.
An iteration's communication volume is therefore fixed by the replicas its placement adds, no matter how the transfers are realized.

Volume alone does not determine completion time, because no realization of a collective finishes before its bottleneck device does.
Completion time is at least the largest per-device send or receive volume divided by that device's link bandwidth $\bw$, so a device that must receive $r$ chunks needs at least $r \Psi / \bw$ time on any topology.
The placement fixes each device's receive count in \collsag and its send count in \collsrs, while leaving open which devices supply the replicas and where gradients are combined.
These per-device bounds can be approached without a fixed transfer schedule, because any device holding a chunk's current replica can supply it, gradients can be combined into partial sums en route, and a chunk is large enough to stream in pieces.
Extra replicas of a chunk thus add bytes to the collective, not proportionally more time at any single device, so the time to materialize and reduce a chunk is insensitive to its replica count.

The same accounting extends from devices to nodes.
When devices group into nodes whose inter-node links are far slower than the intra-node ones, a chunk starts on the node of its shard, and added replicas there are filled over intra-node links alone.
Reaching a node that does not already hold the chunk requires at least one inter-node transfer.
A well-routed transfer delivers it exactly once and broadcasts within the node, so further replicas on a node the chunk has already entered cost no additional inter-node bytes.
Gradient return mirrors this, with one combined partial sum leaving each node the chunk entered.

\textbf{Comparison with FSDP.}
\collsag delivers one complete chunk per pair in $\Delta\set{P}$ and \collsrs returns one gradient chunk per pair, so the two collectives move the same communication volume:
\begin{equation}\label{eq:vol_spcoll}
    \text{Vol}(\collsagop{\set{P}}{\set{P}'})
    = \text{Vol}(\collsrsop{\set{P}'}{\set{P}})
    = |\Delta\set{P}| \cdot \Psi.
\end{equation}
A chunk materialized on $k$ added devices thus contributes $k$ pairs.
The analysis charges one \collsag and one \collsrs per MoE layer per iteration; re-materialization (\refsec{sec:design:model}) adds one more \collsag of the same volume.

FSDP applied to the whole MoE layer is the extreme case $\set{P}' = \set{C} \times \set{D}$, materializing every chunk on every device at every iteration.
For a message of size $s$ over $n$ devices, bandwidth-optimal \collag and \collrs move $(n-1) \cdot s / n$ bytes per device, for a total of $(n-1) \cdot s$ per collective and $2(n-1) \cdot s$ for a full \collar (\collrs followed by \collag)~\cite{chan2007collective}.
FSDP's dense \collag moves $(|\set{D}|-1) \cdot S$ and its \collrs returns the same volume, regardless of token routing.
The sparse collectives' volume is thus the fraction $|\Delta\set{P}| \,/ \left(|\set{C}|(|\set{D}|-1)\right)$ of dense FSDP's, set by the chosen placement.

\textbf{Comparison with Selective Expert-Wise FSDP.}
A different sparsification of FSDP's dense collectives keeps each expert's parameters and optimizer states sliced across all devices~\cite{zhao2023pytorchfsdp}, so every device holds a $1/|\set{D}|$ slice of every expert.
Each device gathers complete weights only for the experts it executes under the iteration's placement, one gather per expert.
Because each optimizer step leaves updated weights only in the slices, no device durably holds a complete expert.
Every gather therefore assembles its expert in full from slices each iteration, delivering the $(1-1/|\set{D}|)\,\Psi$ bytes that the device's own slice does not already supply, even for an expert that keeps executing at the same device.

Under \yyy, the optimizer step instead updates each shard in place, so added replicas are materialized each iteration and weight movement is charged to the placement delta (\refeq{eq:vol_spcoll}).
An expert that needs no added replica executes in place and moves no weight bytes under \yyy, while selective expert-wise FSDP still assembles that same expert in full at whichever device executes it.
More generally, when an expert executes at its shard and at $k$ added devices, the shard is already in place, so \yyy moves $k \Psi$ while selective expert-wise FSDP assembles all $k+1$ replicas for $(k+1)(1-1/|\set{D}|)\,\Psi$.
Selective expert-wise FSDP thus pays $\Psi \left(1-(k+1)/|\set{D}|\right)$ more per expert, a gap that is never negative and closes only at $k+1 = |\set{D}|$, where both sides coincide with dense FSDP.
The gap arises because selective expert-wise FSDP saves only the $1/|\set{D}|$ slice already local at each destination, at most one chunk in total per expert, while \yyy saves the complete shard for every expert.
\yyy sparsifies state and communication alike at exactly the granularity at which the model is sparse; selective expert-wise FSDP sparsifies communication alone.

In aggregate, selective expert-wise FSDP pays for every replica in the placement and \yyy only for the added ones.
As $\Delta\set{P}$ shrinks, selective expert-wise FSDP still assembles the entire placement at a cost that never falls below $(1-1/|\set{D}|)\,S$, while $|\Delta\set{P}| \cdot \Psi$ falls toward zero.
The same contrast holds for completion time.
Gathering from slices spreads the send volume evenly across devices, but each device must still receive $(1-1/|\set{D}|)\,\Psi$ bytes for every expert it executes, even when each expert executes at a single device.
Under \yyy, a device need only receive the replicas the placement adds to it, and nothing when the placement adds none.

\begin{sloppypar}
\textbf{Comparison with Rearrangement.}
Rearrangement-based systems that keep an expert replicated across a device group $\set{D}_i \subseteq \set{D}$ synchronize replica gradients every iteration with one \collar per replicated expert solely to keep the replicas consistent, moving
\begin{equation}\label{eq:vol_ars}
    \text{Vol}(\texttt{{\collar}s}) =
    \sum_{i \,:\, |\set{D}_i| > 1} 2\,(|\set{D}_i| - 1) \cdot \Psi.
\end{equation}
When $\set{P}'$ places each replicated expert on the same group, $\set{P} \subseteq \set{P}'$ puts the expert's shard inside the group and leaves the other $|\set{D}_i| - 1$ devices holding added replicas, so \collsagop{\set{P}}{\set{P}'} and \collsrsop{\set{P}'}{\set{P}} together charge the same per-expert volume by \refeq{eq:vol_spcoll}.
\yyy keeps only the shards and materializes replicas each iteration, so the same communication budget realizes a newly chosen $\set{P}'$ rather than maintaining a fixed placement.
When the placement changes across iterations, those systems additionally migrate the persistent replicas and their optimizer states; \yyy materializes a different $\Delta\set{P}$ without migration.
\end{sloppypar}
\section{\xxx System}\label{sec:arch}

\subsection{Architecture Overview}\label{sec:arch:overview}

\begin{figure}[tb]
    \centering
    \includegraphics[width=0.9\linewidth]{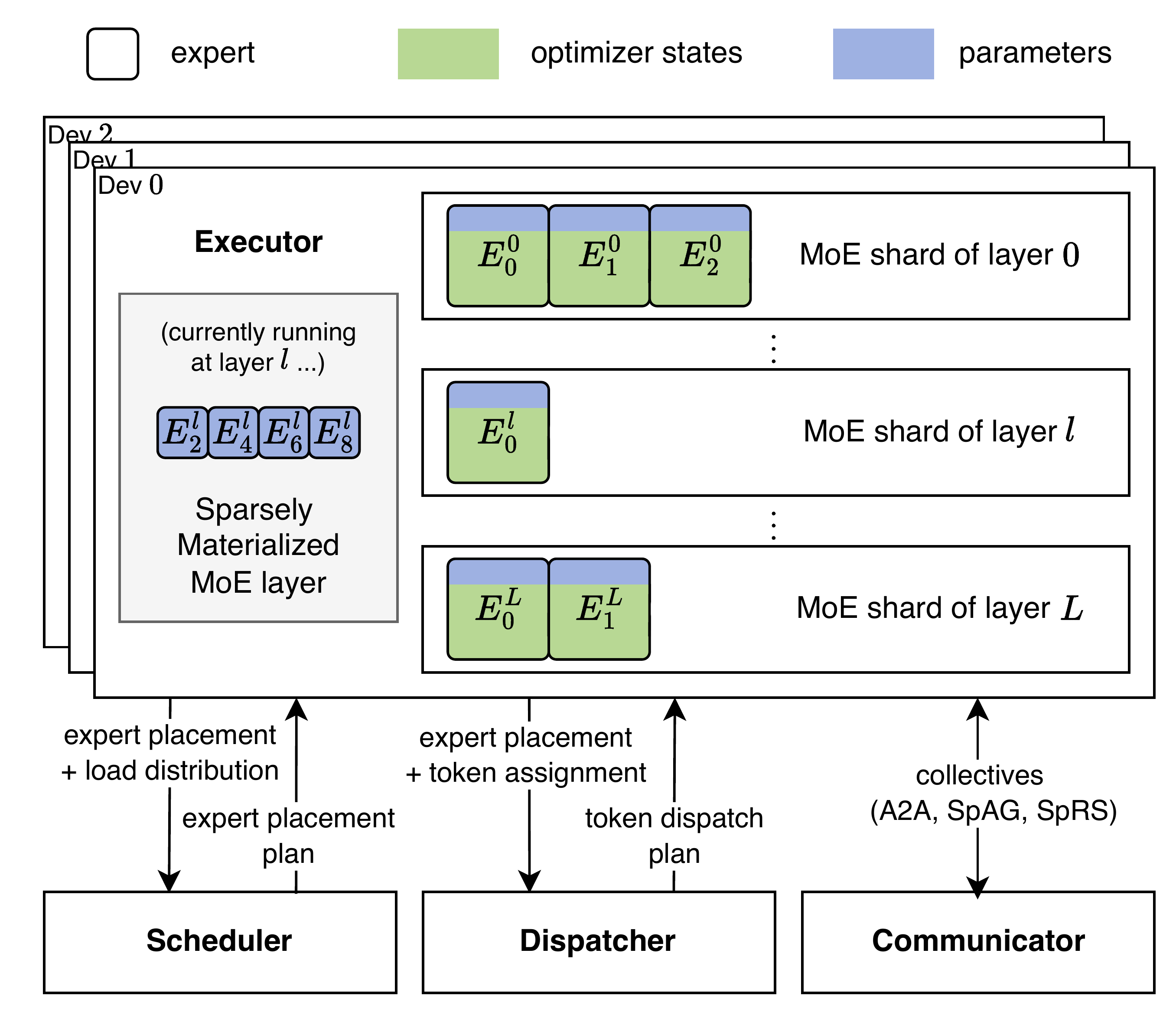}

    \caption{
        \xxx's architecture. 
        The snapshot shows the materialization phase at layer $l$, where each device holds one MoE shard per layer plus the sparsely materialized parameters of layer $l$.
    }

    \label{fig:architecture}
\end{figure}

\yyy fixes how expert state is laid out and moved, but leaves three decisions open: which experts each device owns (the shard placement $\set{P}$), which replicas each iteration's placement $\set{P}'$ adds beyond those shards, and which device in $\set{P}'$ serves each token.
\xxx makes these decisions at runtime.
\reffig{fig:architecture} illustrates the \xxx architecture, where each device launches a runtime consisting of an executor, scheduler, dispatcher, and communicator.
The executor, the main process for each device, controls the \yyy workflow.
In the sharding phase, it queries the scheduler for an MoE layer sharding plan and hosts MoE shards with their expert parameters and optimizer states on the device.
In the materialization phase, it invokes the scheduler for a sparse materialization plan and issues the resulting collectives through the communicator.
When MoE gates assign tokens to experts, the executor queries the dispatcher for each token's target device, since experts may reside on multiple devices.

The scheduler generates expert placement plans based on expert loads to mitigate straggler effects.
It implements two topology-aware algorithms, heterogeneous sharding and sparse materialization, for the two phases of \yyy, respectively.
The dispatcher determines each token's target device based on MoE gate assignments and current placement.
The communicator handles assigned communication tasks by maintaining a queue and scheduling them to the runtime communication library (\eg, NCCL~\cite{nccl2020docs}); it executes the dispatching plan as an \collatoa collective.

The three runtime decisions answer the same bottleneck.
MoE's \collatoa traffic converges on the ingress side, where a device's sends fan out across many receivers but its receives must absorb every token routed to its local experts.
The inter-node links that carry cross-node tokens (\eg NICs~\cite{nvbandwidth2024}) are far slower than the intra-node links (\eg NVLinks~\cite{nvlink2022docs}).
\xxx therefore acts on three time scales: sparse materialization gives overloaded experts replicas on more nodes each iteration so that more tokens find intra-node replicas; heterogeneous sharding periodically spreads underloaded experts' shards to balance per-node ingress load; and token dispatching routes each token to an intra-node device holding the expert when one exists.
Together they reduce inter-node \collatoa traffic and mitigate communication hotspots.

\subsection{Sparse Materialization}\label{sec:arch:sparse_materialization}

In the materialization phase, \yyy requires a sparse materialization plan $\set{P}'$ as the target placement for \collsagop{\set{P}}{\set{P}'}.
Here $\setE = \{E_1, E_2, \ldots\}$ denotes the experts of an MoE layer, each expert's parameters forming one chunk of the sparse collectives.
The scheduler starts from the shard placement $\set{P}$ produced in the sharding phase and extends it with ephemeral replicas, so the plan satisfies $\set{P}' \supseteq \set{P}$.

An ephemeral placement specifies the experts to replicate, the number of replicas for each, and the nodes those replicas cover.
Replica counts follow the expert load distribution, whose skew produces the stragglers to be mitigated, since a replica balances compute regardless of which device holds it.
The nodes the replicas cover determine how much token traffic crosses nodes, because an overloaded expert draws tokens from across the cluster and only a replica on a node that lacks the expert keeps that node's tokens intra-node.
The placement decisions rest on expert loads known only once the MoE gate has run, so the scheduler exploits the temporal locality of the gate's token assignments.
As the gate is trained with stochastic gradient descent~\cite{bottou2010large}, its per-step updates are small, so the load distribution varies little between adjacent iterations~\cite{nie2023flexmoe} even as these shifts accumulate substantially over longer horizons.
The scheduler can therefore predict each iteration's load from recent history and schedule the sparse collectives before MoE computation begins.

\xxx issues each MoE layer's \collsag during the forward computation of the preceding Attention layer~\cite{vaswani2017attention}.
In the backward pass of an Attention layer, two sparse collectives can be overlapped with its computation,
\collsag for re-materializing the MoE parameters of the following layer for gradient computation and
\collsrs for gradient reduction.
Since backward typically takes roughly twice the compute time of the forward pass~\cite{kaplan2020scaling}, constraining \collsag to Attention forward's latency fits both backward sparse collectives within Attention backward's compute window, as illustrated in \reffig[c]{fig:overlap}.

\begin{algorithm}[b!]
    \caption{Sparse Materialization}
    \label{algo:sparse_materialization}
    \SetAlgoLined

    \KwIn{%
        \begin{tabular}[t]{@{}l@{}}
            $\set{P}$: shard placement\\
            $F$: expert load distribution\\
            $t$: overlap degree\\
            $m$: materialization budget per device\\
        \end{tabular}%
    }
    \KwOut{$\set{P}'$: materialization plan}
    
    $t \gets \min(t, |\set{E}|)$, $m \gets \min(m, t)$\;
    $\set{P}' \gets \set{P}$\;
    $\set{E}^\text{topT} \gets$ Top $t$ experts by load $F$\;
    \eIf{$t \leq m$}{
        $\set{P}' \gets \set{P}' \cup (\set{E}^\text{topT} \times \set{D})$\; \label{algo:sparse_materialization:br1start}
    }{ \label{algo:sparse_materialization:br2start}
        $totSlots \gets |\set{D}| \cdot m$\;  
        \ForEach{$e \in \operatorname{sortByLoadDescending}(\set{E}^\text{topT})$}{
            $n \gets \operatorname{assignSlotsByLoad}(e, totSlots, F)$\; \label{algo:sparse_materialization:slot_assignment}
            $\set{P}^{e} \gets$ Distribute $n$ replicas of expert $e$ across nodes and devices, prioritizing nodes that do not already hold a replica of $e$, then devices by available slots\; \label{algo:sparse_materialization:topo_awareness}
            $\set{P}' \gets \set{P}' \cup \set{P}^{e}$\;
        } \label{algo:sparse_materialization:br2end}
    }
    \Return{$\set{P}'$}
\end{algorithm} 
\refalgo{algo:sparse_materialization} presents the topology-aware \textit{sparse materialization} algorithm, based on Longest-Processing-Time-first (LPT) scheduling~\cite{tanaev2012scheduling}.
The algorithm searches for a balanced placement under two system constraints: materialization budget $m$ and overlap degree $t$.
The materialization budget denotes the number of ephemeral expert replicas that fit in each device's spare memory.
Under re-materialization (\refsec{sec:design:model}), each layer's budget spans spare memory that persistent replicas would split across all layers.
The overlap degree $t$ caps the number of distinct experts materialized per layer each iteration so that the sparse collectives fit within the Attention overlap window.
In the cost model of \refsec{sec:design:cost}, a link of bandwidth $\bw$ needs at least $\Psi/\bw$ time per received chunk, where $\Psi$ is one expert's parameter byte size.
This overlap window is $T_\text{attn}$, the forward computation latency of the preceding Attention layer.
Dividing this window by the per-chunk transfer time yields $t = \lfloor T_\text{attn} \cdot \frac{\bw}{\Psi} \rfloor$, the number of expert chunks a link can absorb within it.
A device holds at most one replica of an expert, and a well-routed transfer delivers an expert to each newly reached node once and broadcasts within it, so capping the materialized experts at $t$ holds every receiving link to at most $t$ chunks.
When inter-node bandwidth falls far below intra-node bandwidth, this per-node ingress binds and $\bw$ represents the inter-node bandwidth.
On a homogeneous interconnect, each device's own link binds instead, with $\bw$ the uniform inter-device bandwidth.
Since the computation pattern of an Attention layer is static during training, $T_\text{attn}$ can be profiled before training or captured at runtime.
Expert load $F$ is estimated using a sliding window average over the latest $w$ iterations (\xxx uses $w=5$).

The two outermost branches in \refalgo{algo:sparse_materialization} correspond to which of the two constraints binds.
When the overlap degree is less than or equal to the materialization budget (line~\ref{algo:sparse_materialization:br1start}), the algorithm materializes as many overloaded experts as the overlap window allows, placing each on every device.
Otherwise (lines~\ref{algo:sparse_materialization:br2start} to \ref{algo:sparse_materialization:br2end}), it sparsely materializes experts on devices according to the experts' load distribution.
Experts with higher loads are materialized on more devices (line~\ref{algo:sparse_materialization:slot_assignment}), prioritizing nodes that do not already have the expert parameters materialized (line~\ref{algo:sparse_materialization:topo_awareness}).

The sparse materialization can include a \textit{calibration} stage after MoE gates generate token assignments.
Since overlapped materialization relies on estimated loads, actual distributions may vary due to training stochasticity.
Calibration re-runs \refalgo{algo:sparse_materialization} with the actual loads and the remaining materialization budget, then merges the result with the already-materialized placement.
If the estimated latency reduction exceeds the additional communication cost and that cost fits within a profiled latency budget, \xxx adopts the merged placement and completes the layer's \collsag toward it before token dispatching.

\subsection{Heterogeneous Sharding}\label{sec:arch:hetero_sharding}

Sparse materialization primarily benefits overloaded experts, whose replicas are spread across multiple devices.
The placement of underloaded experts also affects straggler latency, particularly in multi-node training.
Since sparse materialization adds no replicas of underloaded experts, a node containing only such experts becomes the sole ingress point for their tokens, risking bandwidth oversubscription even though each individual expert receives few tokens.

\begin{figure}[b]
    \centering
    \includegraphics[width=\linewidth]{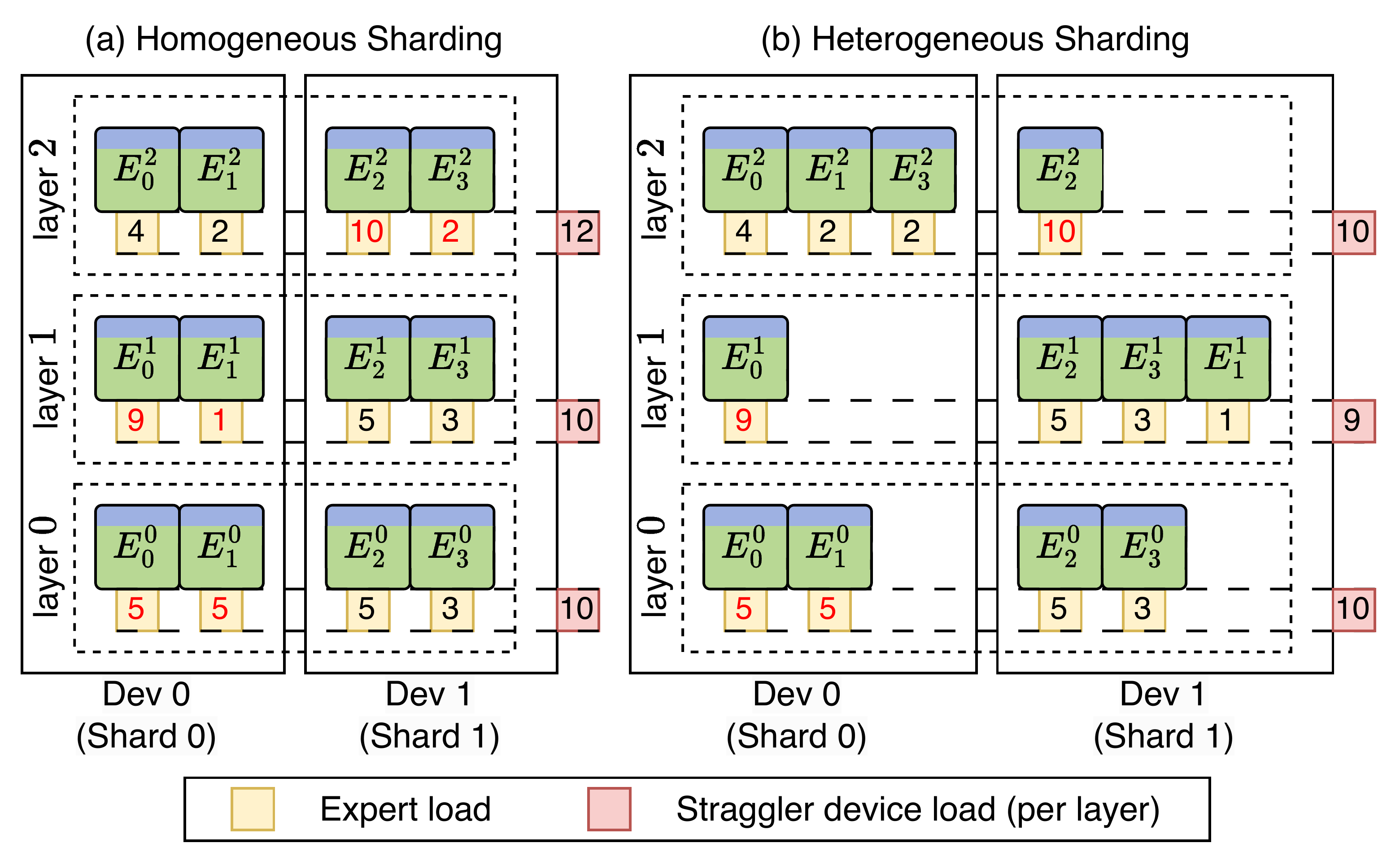}

    \caption{
        Homogeneous vs. heterogeneous sharding across MoE layers.
        Heterogeneous sharding lowers straggler load at the same per-device memory.
    }

    \label{fig:hetero_sharding}
\end{figure}

{\emergencystretch=1em
\xxx's \textit{heterogeneous sharding} algorithm optimizes placement across devices during the sharding phase, targeting underloaded experts in particular.
The algorithm is \textit{heterogeneous} since it allows an MoE shard to have an arbitrary number of experts (ranging from $0$ to $|\set{E}|$) while maintaining memory balance across devices (\reffig{fig:hetero_sharding}).
In \xxx, MoE layers are initialized using homogeneous sharding (\ie even sharding), and periodically re-sharded in a heterogeneous manner.
\par}

\begin{algorithm}[t!]
    \caption{Heterogeneous Sharding}
    \label{algo:hetero_sharding}
    \SetAlgoLined

    \KwIn{%
        \begin{tabular}[t]{@{}l@{}}
            $F^\text{g}$: expert load distribution of all MoE layers \\
            $t$: overlap degree\\
        \end{tabular}%
    }
    \KwOut{$\gset{P}$: MoE sharding plan}

    $\set{J} \gets$ top-$t$ experts by load for each layer \; \label{algo:hetero_sharding:partition-start}
    $\set{J}' \gets \gset{E} - \set{J}$ \; \label{algo:hetero_sharding:partition-end}
    $S_{d} \gets |\gset{E}| / |\set{D}|,\ \forall d \in \set{D}$ ; \tcp{Available slots per device} \label{algo:hetero_sharding:slots}
    $\gset{P} \gets \varnothing$ \;

    \CommentSty{/* Place underloaded experts first. */} \\
    $\set{L} \gets \{\set{E}_{l} \cap \set{J}' \ |\ l = 1, 2, \cdots, L\}$ \; \label{algo:hetero_sharding:underloaded-start}
    \ForEach{$\set{E}'_{l} \in \operatorname{sortByMaxLoadDescending}(\set{L} )$} {
        $\set{P}_l \gets \varnothing$ \;
        \ForEach{$e \in \operatorname{sortByLoadDescending}(\set{E}'_{l})$} {
            $n \gets$ least-loaded node with available slots, prioritizing nodes with fewer available slots \;
            $d \gets$ least-loaded device with $S_{d} > 0$ on node $n$, prioritizing devices with fewer available slots \;
            $\set{P}_l \gets \set{P}_l \cup \{(e, d)\}$\;
            $S_{d} \gets S_{d} - 1$ \;
        }
        $\gset{P} \gets \gset{P} \cup \{\set{P}_l\}$ \;
    } \label{algo:hetero_sharding:underloaded-end}

    \CommentSty{/* Place overloaded experts next. */} \\
    update $\gset{P}$ by arbitrarily placing $\set{J}$ into the remaining slots $S_{d}$ \; \label{algo:hetero_sharding:overlappable}

    \Return{$\gset{P}$}
\end{algorithm}

\refalgo{algo:hetero_sharding} presents the sharding algorithm, which collectively schedules all MoE layers to ensure even memory demand across devices.
Since the algorithm involves cross-layer scheduling, variables use superscript ``g'' for all MoE layers (\eg $\gset{E}$) and index ``$l$'' for specific layers (\eg $\set{E}_l$).
It returns the sharding plan for the model's $L$ MoE layers in the form of $\gset{P} = \{\set{P}_1, \set{P}_2, \cdots, \set{P}_{L}\}$, where each element is an expert placement for the parameters and optimizer states of the corresponding MoE layer.

{\emergencystretch=1em
Experts are partitioned into two disjoint sets layer-wise (lines~\ref{algo:hetero_sharding:partition-start}-\ref{algo:hetero_sharding:partition-end}): $\set{J}$ contains overloaded experts selectable by sparse materialization, and $\set{J}'$ contains the remaining experts.
The algorithm initializes the same number of slots per device (line~\ref{algo:hetero_sharding:slots}) for assigning experts while ensuring consistent memory demand across devices.
Experts in $\set{J}'$ are scheduled first, layer by layer (from line~\ref{algo:hetero_sharding:underloaded-start} to line~\ref{algo:hetero_sharding:underloaded-end}), processing layers in descending order of the maximum per-expert load within each layer's $\set{J}'$ subset.
Each expert is assigned to the least-loaded node and, within that node, to the least-loaded device, considering only nodes and devices with remaining slots; ties at both levels are broken by fewest remaining slots.
Placing an expert consumes one slot on its device.
Finally, the algorithm fills the remaining slots with experts from $\set{J}$ (line~\ref{algo:hetero_sharding:overlappable}).
\par}

{\emergencystretch=1em
Unlike sparse materialization, whose collectives are scheduled into computation overlap windows, heterogeneous sharding introduces re-sharding latency directly to the critical path.
Re-sharding, however, need only happen infrequently.
The targeted underloaded experts receive fewer tokens per iteration, and fewer tokens induce smaller gradient magnitudes on the gate, so their loads drift slowly across iterations, as their traces in \reffig{fig:trace} show.
The slow drift allows \xxx to amortize re-sharding overhead over many iterations while retaining \yyy's per-iteration load-balancing gains.
\par}

\subsection{Token Dispatching}\label{sec:arch:dispatch}

With sparse materialization, an expert's parameters may exist on multiple devices.
Each token assigned to this expert must be routed to one of the devices holding its parameters.
\xxx's dispatcher generates a topology-aware token dispatching plan, presented in \refalgo{algo:token_dispatching}.
The algorithm aims to minimize inter-node communication.
It takes as input the token assignment $T$, where $T_{e,d}$ is the number of tokens from device $d$ assigned to expert $e$ by the MoE gate, and the materialized parameter placement $\set{P}'$; it outputs a dispatching plan $R$, where $R_{e,d,d'}$ denotes the number of tokens on device $d$ to be dispatched to expert $e$ materialized on device $d'$.

\begin{algorithm}[!b]
    \caption{Topology-aware Token Dispatching}
    \label{algo:token_dispatching}
    \SetAlgoLined

    \KwIn{%
        \begin{tabular}[t]{@{}l@{}}
            $\set{P}'$: materialized parameter placement \\
            $T$: token assignment from MoE gate \\
        \end{tabular}%
    }
    \KwOut{$R$: dispatching plan}

    Initialize $R$ \; 
    \ForEach{$e \in \set{E}$}{
        $\set{D}^e \gets$ devices having expert $e$ materialized \;
        \ForEach{$d \in \set{D}^e$}{ \label{algo:token_dispatching:local-start}
            \CommentSty{/* Local dispatching. */}
            $R_{e,d,d} \gets T_{e,d}$ \;
        } \label{algo:token_dispatching:local-end}
        \ForEach{$d \in (\set{D} - \set{D}^e)$}{ \label{algo:token_dispatching:nonlocal-start}
            $\set{D}^{local} \gets$ all devices in the same node as $d$ \;
            $\set{D}^{elocal} \gets \set{D}^{e} \cap \set{D}^{local}$ \;
            \eIf{$\set{D}^{elocal} \ne \varnothing$}{
                \CommentSty{/* Intra-node dispatching. */}
                $R_{e,d,:} \gets $ evenly dispatch $T_{e,d}$ to $\set{D}^{elocal}$ \;
            }{
                \CommentSty{/* Inter-node dispatching */}
                $R_{e,d,:} \gets $ evenly dispatch $T_{e,d}$ to $\set{D}^e$ \;
            }
        } \label{algo:token_dispatching:nonlocal-end}
    }
    \Return{$R$}
\end{algorithm}
 
The algorithm operates in two phases for each expert $e$, over the devices $\set{D}^e$ holding its parameters.
First, tokens on devices where the expert is already materialized are dispatched locally (lines~\ref{algo:token_dispatching:local-start}--\ref{algo:token_dispatching:local-end}), incurring zero communication cost.
Second, for devices without the expert locally (lines~\ref{algo:token_dispatching:nonlocal-start}--\ref{algo:token_dispatching:nonlocal-end}), the algorithm first tries intra-node devices that already hold the expert and otherwise falls back to all holders, distributing tokens evenly in either case.
The local-first preference keeps token traffic intra-node where possible and reduces contention on inter-node links.
\section{Evaluation}\label{sec:eval}

\subsection{Experimental Setup}\label{sec:eval:setup}

\begin{figure*}[!ht]
    \centering
    \includegraphics[width=\linewidth]{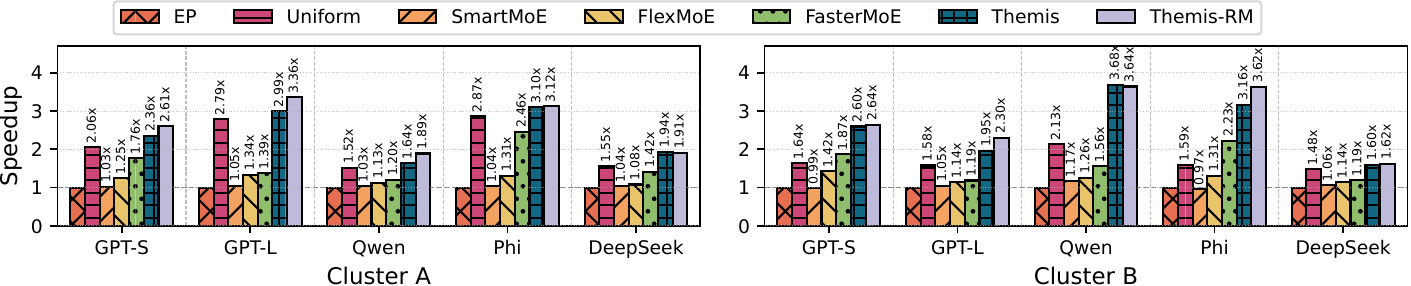}
    \caption{End-to-end MoE training speedups over EP on diverse MoE models and cluster configurations.}
    \label{fig:end2end}
\end{figure*} 
\textbf{Implementation.}
\xxx's sparse collectives are prototyped with NCCL~\cite{nccl2020docs} group calls to simultaneously schedule a series of \collbc and \collrd operations.
While optimized algorithms could further exploit the sparse data distribution and the network topology, our prototype moves no more than the communication volume derived in \refsec{sec:design:cost}. We leave optimized implementation to future work.

\begin{sloppypar}
\textbf{Testbeds.}
We evaluate \xxx on two cloud clusters:
\clustera with AWS p4de.24xlarge nodes, each containing 8 NVIDIA A100-80G GPUs interconnected via 600\,GB/s NVSwitch~\cite{nvswitch2022docs}, with 400\,Gbps inter-node network;
\clusterh with AWS p5en.48xlarge nodes, each containing 8 NVIDIA H200-141G GPUs interconnected via 900\,GB/s NVSwitch~\cite{nvswitch2022docs}, with 3200\,Gbps inter-node network.
Unless otherwise noted, experiments use 4 nodes (32 GPUs).
Our software environment is based on the NGC PyTorch 24.12 container, with CUDA 12.6, NCCL 2.23.4~\cite{nccl2020docs}, and PyTorch 2.6~\cite{paszke2019pytorch}.
\end{sloppypar}

\textbf{Baselines.}
We compare \xxx against three baseline systems.
All compared systems, including \xxx, run within the same MoE training framework integrated with Megatron-LM~\cite{shoeybi2019megatron} and differ only in their expert placement and scheduling policies.
We use DP to parallelize non-MoE layers (e.g., Attention) across GPUs.
FasterMoE~\cite{he2022fastermoe} broadcasts overloaded experts to all devices after gating, guided by a latency model that weighs broadcast cost against load-balancing benefit.
SmartMoE~\cite{zhai2023smartmoe} permutes experts between placement slots to colocate overloaded and underloaded ones.
FlexMoE~\cite{nie2023flexmoe} adjusts persistent replica counts one unit at a time, expanding overloaded experts and shrinking underloaded ones until reaching a balance threshold.
For each set of comparative experiments, we use the largest batch size that avoids OOM across all systems, so every system trains the same model under the same batch configuration.
Unless otherwise noted, \xxx re-shards only once every 100 iterations.
\xxx-RM denotes \xxx with parameter releasing and re-materialization.
``Uniform'' denotes EP with a special gate that dispatches tokens evenly across all experts.
\begin{table}[htbp]
    \centering
    \caption{Sizes and architectures of evaluated MoE models. Layer counts given as a/b denote the \clustera and \clusterh configurations, respectively.}
    \label{tab:models}
    \setlength\tabcolsep{2pt}
    \begin{tabular}{lrrrccc}
        \toprule
        Model & $d_\text{model}$ & $d_\text{ffn}$ & SeqLen & Layers & E/GPU & Top-$k$ \\
        \midrule
        GPT-S-MoE & 768 & 3072 & 2048 & 12 & 2 & 2 \\
        GPT-L-MoE & 1536 & 6144 & 2048 & 12 & 2 & 2 \\
        \mphi & 4096 & 6400 & 4096 & 4/8 & 1 & 2 \\
        \mqwen & 2048 & 1408 & 4096 & 12/24 & 2 & 4 \\
        \mds & 7168 & 2048 & 4096 & 2/4 & 4 & 8 \\
        \bottomrule
    \end{tabular}
\end{table}

\textbf{Models and Metrics.}
We evaluate \xxx on both synthetic and production MoE models, as detailed in Table~\ref{tab:models}.
For synthetic models, following prior work~\cite{he2022fastermoe,zhai2023smartmoe,nie2023flexmoe}, we sparsify GPT-3~\cite{brown2020language} by replacing its FFNs with MoE layers, where each expert is an FFN with the original model dimension ($d_\text{model}$) and hidden dimension $d_\text{ffn} = 4 \times d_\text{model}$.
To demonstrate \xxx's scalability on production workloads, we further evaluate three industrial open-source MoE models: \mphi~\cite{abdin2024phi3technicalreporthighly}, \mqwen~\cite{bai2023qwentechnicalreport}, and \mds~\cite{deepseekai2025deepseekv3technicalreport}.
For \mds, we use the same E/GPU ratio as the original paper~\cite{deepseekai2025deepseekv3technicalreport} and evaluate a 4-layer model on \clusterh (2-layer on \clustera due to memory constraints), matching the per-stage layer count of the original 61-layer model with 16-way pipeline parallelism.
Standard training-time load-balancing mechanisms are enabled, with all models using an auxiliary load-balancing loss (weight 0.001) unless otherwise specified.
We do not use token dropping techniques such as expert capacity~\cite{lepikhin2020gshard}, which degrade statistical efficiency by discarding informative tokens~\cite{nie2023flexmoe,gale2023megablocks} and have been abandoned by recent production MoE systems~\cite{deepseekai2025deepseekv3technicalreport,yang2025qwen3technicalreport}.
Varying SeqLen tests \xxx under different overlap windows between sparse collective communication and Attention computation.

\subsection{End-to-End Performance}\label{sec:eval:end2end}

We report end-to-end throughput relative to EP in \reffig{fig:end2end}.
\xxx achieves up to \maxSpeedupEP speedup over EP, with average speedups of \avgSpeedupA on \clustera and \avgSpeedupH on \clusterh.

Among rearrangement-based systems, FasterMoE consistently outperforms SmartMoE and FlexMoE across all configurations.
Its latency model triggers a rearrangement only when the predicted load-balancing benefit outweighs the transfer cost.
SmartMoE and FlexMoE instead adjust placements at fixed intervals or when imbalance crosses a threshold, without weighing each adjustment's cost against its benefit. The adjustment cost and the lag behind shifting loads occasionally make them slower than vanilla EP in our measurements.
\xxx outperforms FasterMoE by up to \maxSpeedupFasterMoE (\minSpeedupFasterMoE to \maxSpeedupFasterMoE across workloads), and achieves even larger gains over SmartMoE (up to \maxSpeedupSmartMoE) and FlexMoE (up to \maxSpeedupFlexMoE).
The speedup reflects three factors: overlapping sparse materialization with Attention computation, topology-aware placement that reduces inter-node traffic, and heterogeneous sharding that mitigates residual stragglers.
\xxx-RM's released replicas raise its per-layer materialization budget (\refsec{sec:arch:sparse_materialization}), and the better balance the larger budget admits outweighs the additional sparse collective invocations, so \xxx-RM outpaces \xxx in most configurations and outperforms all baselines.

Uniform dispatch eliminates compute skew and provides a reference for imbalance overhead, where EP is \imbalanceOverheadMin to \imbalanceOverheadMax slower than Uniform due to stragglers.
\xxx and \xxx-RM not only close this gap in all cases but surpass Uniform by \speedupUniformMin to \speedupUniformMax, as topology-aware placement reduces inter-node traffic beyond what balanced routing alone achieves.

\subsection{Fine-Grained Performance Breakdown}\label{sec:eval:breakdown}

\begin{figure}[tb]
    \centering
    \includegraphics[width=\linewidth]{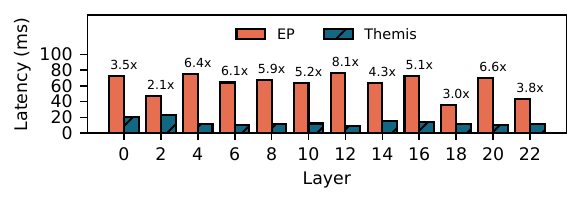}
    \caption{Layer-wise speedup of \xxx over EP when training \mqwen on \clusterh.}
    \label{fig:layerwise}
\end{figure} 
\reffig{fig:layerwise} illustrates the layer-wise speedup of \xxx when training \mqwen on \clusterh.
\xxx consistently outperforms EP across all layers, yielding a \layerwiseSpeedupRange speedup.
Load imbalance varies across layers, producing different per-layer execution times under EP.
A system that reserves identical memory for placement in every layer (e.g., FlexMoE) over-provisions some layers and starves others.
Heterogeneous sharding redistributes materialization budgets across layers to match each layer's imbalance, with no net memory increase.

\begin{figure}[tb]
    \centering
    \includegraphics[width=\linewidth]{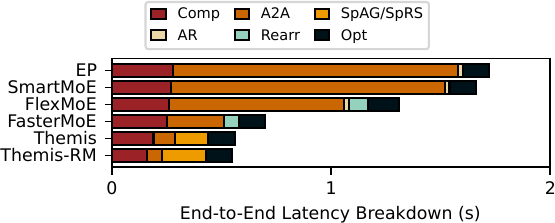}
    \caption{Breakdown of the per-iteration critical path for \mphi on \clustera, accumulated on a representative device. SpAG/SpRS and AR bars show only the time left exposed after overlap; Rearr covers each system's rearrangement transfers.}
    \label{fig:perf_breakdown}
\end{figure} 
\reffig{fig:perf_breakdown} breaks down per-iteration training time of baseline systems and \xxx for \mphi on \clustera, accumulated on a representative device.
A2A covers token \collatoa, in which straggler-induced waiting also surfaces; Rearr covers rearrangement transfers; Opt covers the optimizer step; and Comp covers the forward and backward computation.
Bars accumulate critical-path time, and no system overlaps \collatoa with computation.
SpAG/SpRS and AR bars show the time left exposed after \xxx overlaps sparse collectives with Attention and the non-MoE layers' data-parallel \collar synchronization with the backward pass, respectively.
As illustrated, A2A accounts for the majority of MoE training latency under EP.
Rearrangement-based systems reduce this straggler waiting at the cost of the transfers visible as their Rearr time.
\xxx's topology-aware placement yields the lowest \collatoa time, a \ataReduction reduction over EP.
\xxx-RM incurs additional overhead due to re-materialization, resulting in a \rmOverheadIncrease increase in the sparse collective communication overhead, while still outperforming baseline systems by \rmSpeedupOverBaselines.

Together, these measurements explain where \xxx's speedup comes from.
The gains are the reduced straggler waiting inside \collatoa, the balanced device-level computation, and the avoided rearrangement transfers that appear as the baselines' Rearr time.
The costs are the sparse-collective time left exposed after overlap, the time to compute the materialization plan, and re-sharding latency amortized over its interval (\reffig[b]{fig:component}).
The median time to compute the materialization plan is 24--51\,$\mu$s per MoE layer (at most 67\,$\mu$s at the 95th percentile).
Total expert FLOPs are unchanged, so the gains come from where state and tokens are placed.

\begin{figure}[tb]
    \centering
    \includegraphics[width=\linewidth]{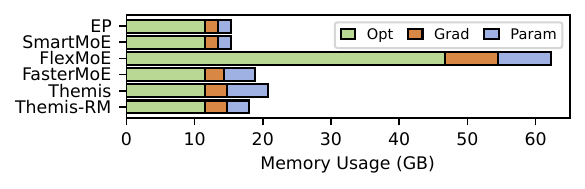}
    \caption{Model state memory breakdown: optimizer states (Opt), gradients (Grad), and parameters (Param). Activation memory is excluded; peak memory including activations is evaluated in \reffig[a]{fig:oom_scale}.}
    \label{fig:mem_breakdown}
\end{figure} 
\begin{figure*}[!ht]
    \centering
    \includegraphics[width=\linewidth]{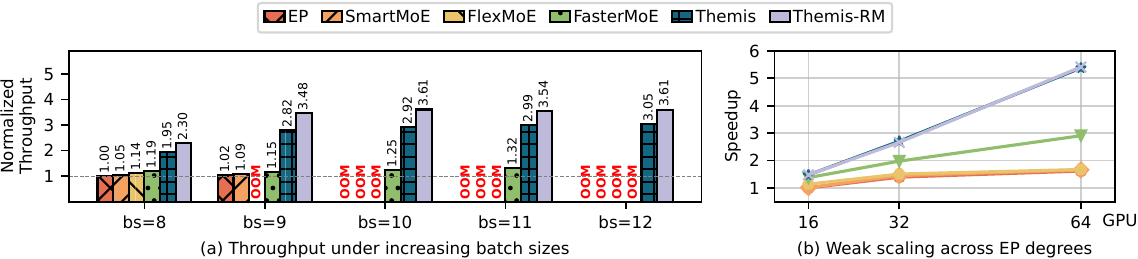}
    \caption{(a) Memory efficiency of GPT-L-MoE on \clusterh under increasing batch sizes.
    (b) Weak scaling of \mds on \clustera with increasing EP degree.}
    \label{fig:oom_scale}
\end{figure*} 
\reffig{fig:mem_breakdown} presents the model state memory breakdown (optimizer states, gradients, and parameters).
FlexMoE uses the most memory, \flexmoeMemOverhead that of \xxx, due to its pre-allocated per-layer reserves.
With sufficient memory, \xxx uses more parameter memory (\themisParamMemVsEP over EP) to materialize the placement that best balances load, a \themisTotalMemIncrease increase in total memory over EP.
\xxx-RM releases the materialized parameters after use, reducing parameter memory by \rmParamMemReduction compared to \xxx, and consequently consumes only \rmTotalMemIncrease more total memory than EP.

\subsection{Ablation Study}

\noindent
\textbf{Memory Efficiency.}
We assess memory efficiency of GPT-L-MoE on \clusterh by incrementally increasing the batch size until systems encounter out-of-memory (OOM) errors.
Memory pressure in MoE training arises from two main sources: (i) overhead from the expert placement strategy and (ii) peak memory consumption from intermediate activations, which is exacerbated by skewed token distribution.
The results in \reffig[a]{fig:oom_scale} reveal a clear hierarchy of memory resilience.
FlexMoE fails first (at bs=9), constrained by its large, pre-allocated memory reserves for rearrangement.
It is followed by systems with less effective load balancing, EP and SmartMoE (at bs=10), and then FasterMoE (at bs=12), all of which suffer from activation memory spikes on overloaded devices.
In contrast, both \xxx and \xxx-RM run at all tested batch sizes up to 12.
For each layer, \xxx-RM frees materialized parameters after the forward pass, reusing the memory across layers.

\begin{figure}[tb]
    \centering
    \includegraphics[width=\linewidth]{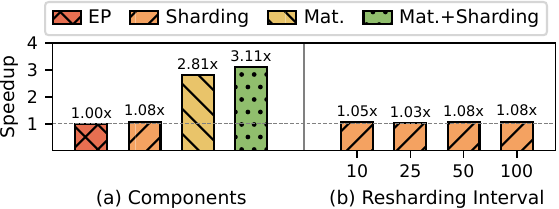}
    \caption{Training \mphi with (a) different combinations of heterogeneous sharding (Sharding) and per-iteration sparse materialization (Mat.);
    (b) heterogeneous sharding under different re-sharding intervals, showing insensitivity to the trigger frequency.}
    \label{fig:component}
\end{figure} 
\noindent
\textbf{Component Effectiveness.} 
We evaluate the effectiveness of \xxx's components (\reffig{fig:component}).
\xxx's heterogeneous sharding achieves consistent speedups (\heteroSpeedupA, \heteroSpeedupB, \heteroSpeedupC and \heteroSpeedupD) across different re-sharding intervals ranging from 10 to 100 iterations.
Heterogeneous sharding is insensitive to the interval, so \xxx re-shards infrequently with negligible overhead and no interval tuning.
Sharding alone (\ie \xxx executing plain EP with a zero materialization budget) rebalances the shard placement only at low frequency, so its standalone gain is modest; its cross-layer memory allocation instead shapes the per-layer materialization budgets exploited every iteration.
The combination is therefore crucial for efficient load balancing, achieving a \matHeteroVsHetero speedup over \xxx with only heterogeneous sharding enabled and a \matHeteroVsMat speedup over \xxx with only materialization enabled.
\begin{figure}[tb]
    \centering
    \includegraphics[width=\linewidth]{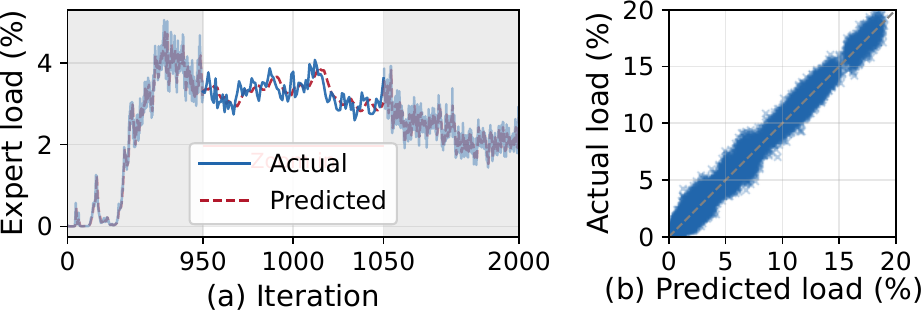}
    \caption{Load prediction of \loadPredModel on \loadPredCluster.
    (a) Predicted vs. actual load for a representative expert over training.
    (b) Prediction accuracy across all experts and MoE layers ($R^2$=\loadPredRSquared).
    }
    \label{fig:load_pred}
\end{figure} 
\noindent
\textbf{Load Prediction Accuracy.}
\xxx relies on load prediction to determine materialization plans before actual token routing is known (\refsec{sec:arch:sparse_materialization}).
Prediction affects only where experts are materialized and where tokens are dispatched, so a misprediction yields a less efficient placement rather than an incorrect result.
\reffig[a]{fig:load_pred} traces the predicted and actual load for a representative expert over training.
Despite per-iteration fluctuations, the predicted curve closely tracks the actual load throughout training, including during high-variance intervals.

\reffig[b]{fig:load_pred} quantifies prediction accuracy across all experts and MoE layers ($R^2$=\loadPredRSquared).
When predictions occasionally deviate due to routing shifts, the calibration stage (\refsec{sec:arch:sparse_materialization}) detects the mismatch after gating and issues an additional materialization transfer only when doing so reduces overall latency.
This two-stage approach preserves the benefits of proactive materialization. The cost of calibration, including the transfers it issues, is counted in the SpAG/SpRS time of \reffig{fig:perf_breakdown}.

\begin{figure}[tb]
    \centering
    \includegraphics[width=\linewidth]{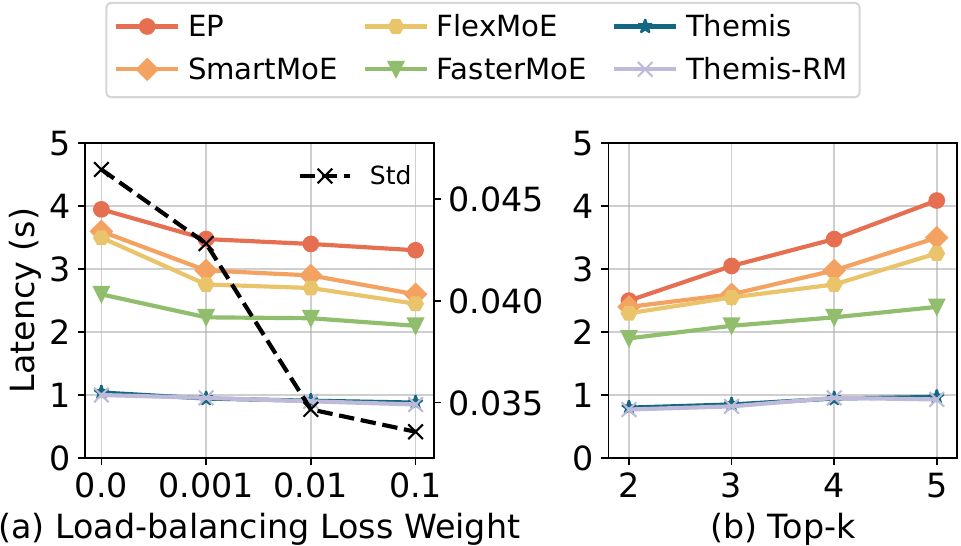}
    \caption{\mqwen on \clusterh with (a) various load-balancing loss weights and (b) varying top-$k$ values. The right axis of (a) shows routing skew as the Std of expert load proportions.}
    \label{fig:sensitivity}
\end{figure} 
\noindent
\textbf{Sensitivity to Load Imbalance.}
We evaluate \xxx's robustness to varying degrees of load imbalance induced by two factors: the auxiliary load-balancing loss weight (\reffig[a]{fig:sensitivity}) and the number of selected experts $k$ in top-$k$ gating (\reffig[b]{fig:sensitivity}).
To quantify residual imbalance, we report the standard deviation (Std) of the per-expert token proportion on the secondary y-axis in \reffig[a]{fig:sensitivity}, computed over all experts with $\text{Std}=0$ indicating perfectly uniform routing.
In \reffig[a]{fig:sensitivity}, increasing the loss weight from 0.0 to 0.1 reduces Std from 0.046 to 0.034, confirming that stronger penalties yield more balanced routing while substantial skew remains.
All systems benefit from reduced skew, with EP, FasterMoE, and FlexMoE reducing latency by 16.5\%, 19.2\%, and 30.0\%, respectively.
In contrast, \xxx and \xxx-RM exhibit the smallest reduction (15.4\%), indicating they already handle high-skew workloads effectively and leave less room for improvement as routing becomes more balanced.
Among baselines, FlexMoE is initially 34.6\% slower than FasterMoE under high skew, but this gap narrows to 16.7\% as routing becomes more balanced. FasterMoE's latency model triggers rearrangement less frequently when imbalance is mild, while FlexMoE benefits from reduced migration overhead when its rearrangement threshold is seldom exceeded.

Increasing $k$ routes each token to more experts, proportionally amplifying \collatoa traffic.
\reffig[b]{fig:sensitivity} shows that all systems experience higher latency as $k$ grows from 2 to 5.
However, \xxx exhibits the flattest growth, with latency increasing by only 21.2\% compared to EP's 63.6\%.
This resilience arises because \xxx materializes overloaded experts on devices where their tokens originate, converting inter-node A2A transfers into local or intra-node communication regardless of how many experts each token activates.

\begin{figure}[tb]
    \centering
    \includegraphics[width=\linewidth]{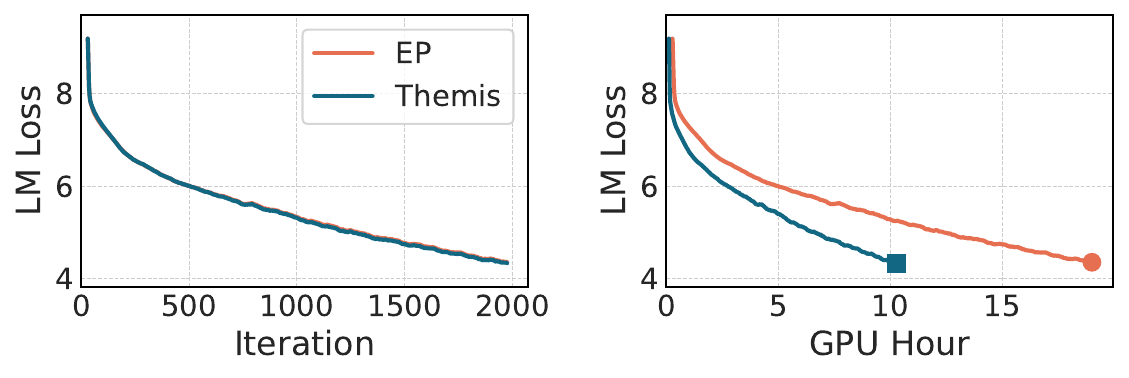}
    \caption{Training loss of \mphi on \clusterh with respect to (a) training steps and (b) GPU hours.}
    \label{fig:loss}
\end{figure} 
\noindent
\textbf{Scalability.}
\reffig[b]{fig:oom_scale} reports weak scaling of \mds along the EP (for experts) and DP (for non-expert modules) dimensions, from 16 to 64 GPUs on \clustera.
While baseline systems suffer increasing latency at larger EP degrees due to worsening load imbalance, \xxx maintains near-constant latency, widening its advantage from \scalabilitySpeedupMin to \scalabilitySpeedupMax over EP.
Since production EP degrees range from 4 (Qwen3~\cite{yang2025qwen3technicalreport}) through 8 (DeepSeek-V2~\cite{deepseekai2024deepseekv2strongeconomicalefficient}) to 64 (DeepSeek-V3~\cite{deepseekai2025deepseekv3technicalreport}), scaling to 64 GPUs covers even the largest EP degree deployed in production.

\noindent
\textbf{Training Convergence.}
\xxx accelerates MoE training without modifying the model architecture or the training algorithm, thereby preserving the convergence characteristics of baseline EP.
To verify this, we train \mphi on \clusterh twice, once with \xxx and once with baseline EP, keeping all hyperparameters and the random seed identical across both runs.
As shown in \reffig[a]{fig:loss}, \xxx's loss curve over training steps is indistinguishable from the baseline's.
Owing to its higher throughput, \xxx reaches the same loss in fewer GPU hours (\reffig[b]{fig:loss}).
\section{Related Work}\label{sec:related}

\textbf{Load Balancing in MoE Training.}
MoE gates commonly use an auxiliary loss~\cite{lepikhin2020gshard} during training to balance tokens across experts, yet load imbalance persists.
Systems like GShard~\cite{lepikhin2020gshard} and Switch Transformer~\cite{fedus2022switch} further limit the expert capacity (i.e., the number of tokens an expert can process) and drop excess tokens for each expert.
However, MegaBlocks~\cite{gale2023megablocks} showed that token dropping degrades model quality and should be avoided.
Routing-side balancing, whether auxiliary losses or capacity limits, is orthogonal to \xxx, which balances expert placement under any load distribution the gate produces.

\textbf{MoE Training Systems.}
DeepSpeed-MoE~\cite{rajbhandari2022deepspeed} establishes GPT-style expert-parallel MoE training at scale, and Tutel~\cite{hwang2022tutel} adds adaptive parallelism and pipelining.
Later systems optimize token \collatoa~\cite{li2023lina,shi2024schemoe}, overlap MoE communication with computation at the scheduler, compiler, or kernel level~\cite{pan2025fsmoe,jiang2024lancet,zhang2025comet}, reuse pipeline memory~\cite{zhang2024mpmoe}, and tolerate faults through adaptive expert placement~\cite{wu2024lazarus}.
Janus~\cite{liu2023janus} fetches expert weights instead of tokens when the weights are the smaller transfer.
FasterMoE~\cite{he2022fastermoe}, SmartMoE~\cite{zhai2023smartmoe}, FlexMoE~\cite{nie2023flexmoe}, and others~\cite{zhang2025popfetcher,wang2023prophet} counter load imbalance by replicating or migrating experts at runtime.
EfficientMoE~\cite{zeng2025efficientmoe} extends such replication to static-graph training on Ascend NPUs.

\textbf{Sparse Collective Communication.}
In distributed deep learning, sparse collectives are often combined with top-$k$ sparsification to reduce gradient communication~\cite{peng2024sparse, zhao2024spardl, li2022near, renggli2019sparcml, shi2019distributed}.
\xxx instead applies sparse collectives to load balancing, moving expert weights rather than sparsified gradients.
Collective-algorithm synthesizers~\cite{cai2021msccl,shah2023taccl,kim2024tccl,wang2020blink,cowan2022gc3} generate data routing plans tailored to specific collective patterns and network topologies.
Such synthesizers could help realize efficient sparse collectives for \yyy.
\section{Conclusion}\label{sec:concl}

\begin{sloppypar}
We presented \yyy, a sparse-native MoE training approach that achieves per-iteration in-situ load balancing through ephemeral expert materialization, removing persistent state migration from the per-iteration critical path.
\yyy introduces sparse collectives for efficient parameter materialization and gradient synchronization, along with heterogeneous sharding and re-materialization for memory-efficient
training.
We realize \yyy in \xxx with topology-aware placement scheduling and token dispatching.
\xxx achieves \overallSpeedup speedup over the state-of-the-art expert-rearrangement systems.
\end{sloppypar} 
\balance
\bibliographystyle{ACM-Reference-Format}
\bibliography{references}

\end{document}